\documentclass[journal]{vgtc}                

\ifpdf
  \pdfoutput=1\relax                   
  \usepackage{graphicx}                
  \DeclareGraphicsExtensions{.pdf,.png,.jpg,.jpeg} 
\else
  \usepackage{graphicx}                
  \DeclareGraphicsExtensions{.eps}     
\fi%

\graphicspath{{figures/}{pictures/}{images/}{./}} 

\usepackage[dvipsnames]{xcolor}

\usepackage{subfig}
\usepackage{tikz-cd}
\usepackage{amsmath}
\usepackage{amsfonts}
\usepackage{bbold}

\usepackage{microtype}                 
\PassOptionsToPackage{warn}{textcomp}  
\usepackage{textcomp}                  
\usepackage{mathptmx}                  
\usepackage{times}                     
\usepackage{cite}                      
\usepackage{tabu}                      
\usepackage{booktabs}                  

\newcommand{\newtext}[1]{#1}

\onlineid{1202}

\vgtccategory{Research}
\vgtcpapertype{Algorithm/technique}

\title{Discriminability Tests for Visualization Effectiveness and Scalability}

\author{Rafael Veras and Christopher Collins}
\authorfooter{
\item
 The authors are with Ontario Tech University. \\E-mail: rafael.verasguimaraes@uoit.ca; christopher.collins@uoit.ca
}

\shortauthortitle{Discriminability Tests for Vis}

\abstract{The scalability of a particular visualization approach is limited by the ability for people to discern differences between plots made with different datasets. Ideally, when the data changes, the visualization changes in perceptible ways. This relation breaks down when there is a mismatch between the encoding and the character of the dataset being viewed. Unfortunately, visualizations are often designed and evaluated without fully exploring how they will respond to a wide variety of datasets. We explore the use of an image similarity measure, the Multi-Scale Structural Similarity Index (MS-SSIM), for testing the discriminability of a data visualization across a variety of datasets. MS-SSIM is able to capture the similarity of two visualizations across multiple scales, including low level granular changes and high level patterns. Significant data changes that are not captured by the MS-SSIM indicate visualizations of low discriminability and effectiveness. The measure's utility is demonstrated with two empirical studies. In the first, we compare human similarity judgments and MS-SSIM scores for a collection of scatterplots. In the second, we compute the discriminability values for a set of basic visualizations and compare them with empirical measurements of effectiveness. In both cases, the analyses show that the computational measure is able to approximate empirical results. Our approach can be used to rank competing encodings on their discriminability and to aid in selecting visualizations for a particular type of data distribution.%
} 


\CCScatlist{ 
 \CCScat{K.6.1}{Management of Computing and Information Systems}%
{Project and People Management}{Life Cycle};
 \CCScat{K.7.m}{The Computing Profession}{Miscellaneous}{Ethics}
}

\vgtcinsertpkg

\begin{document}



\maketitle

\section{Introduction}

One measure of a visualization's effectiveness is whether data changes result in equivalent and perceptible visual changes. That is, can a viewer discriminate between visualizations of different data? Discriminability can be limited by both data scale (e.g., too much data results in overplotting such that views look the same) and by the perceptual scalability of the technique (e.g., small data changes result in changes to hue, position, or other encodings which are too small for a viewer to detect).
Whether a visualization is robustly discriminable across the range of possible datasets is challenging to know at design time and expensive to empirically evaluate. In this work we introduce a process for using the Multi-Scale Structural Similarity Index (MS-SSIM), borrowed from image quality analysis, with simulated data variations, to estimate the discriminability of visual encodings. We evaluate our approach with a crowdsourced experiment comparing human similarity judgments to the MS-SSIM approach, and by using MS-SSIM to approximate previously reported empirical measures of effectiveness.

One predominant process for visualization design and validation is described by Munzner's nested model~\cite{Munzner2009}. This model is structured as nested steps for visualization design and methods for validating each step: a) domain problem and data characterization; b) operation and data type abstraction; c) visual encoding and interaction design; d) algorithm design. At the first level, the designer ``must learn about the tasks and the data of target users in some particular domain''.  Also known as elicitation of requirements, this phase borrows methods from human-centred design, such as ethnographic studies.

We argue that, in practice, this step is conflated into learning about the tasks of the users \textit{in detriment} of the data. The very nested model is a victim of this reduction: the output of step (a) is a ``set of questions asked about or actions carried out by the target users for some heterogeneous data''. Note how the \textit{characterization} of data is not present in the output. In the next level, operation and data type abstraction, the output is a description of operations and data types. Characterizing data is thus reduced to descriptions of data type. This gap gives rise to what we call \textit{exemplary datasets}, a small collection of datasets taken as representative of the population and which the rest of the design process, including evaluation, is based upon. The outcome of the design process is commonly overfit to these few datasets.

The narrow scope of evaluation in the data axis threatens the validity of research claims and the robustness of visualization products. It affects any new encodings or techniques that are expected to be effective over a large range of data. In statistical terms, an exemplary dataset is only a single outcome of the random process that governs the data, and the more dimensions involved, the broader is the data universe. Thus, validity depends on the relation between the tested data and the possible data.



\newtext{To strengthen validity, we can evaluate a visualization against many datasets, either real or produced by simulation.
However, a large collection of test datasets across a large number of participants creates scalability problems for user studies. In this research we propose an automated way to evaluate visualization effectiveness and scalability against large data collections. We contribute a) a procedure for quantifying the discriminability of visual encodings based on a computational measure of image similarity; b) validation of this measure against empirical plot similarity data; and c) validation of the discriminability scores against empirical data on the effectiveness of visual encodings.}

\section{Related Work}

In this section, we position our work in the taxonomy of quality measures proposed by Behrisch et al.~\cite{Behrisch2018}, and use some of its categories to discuss how our discriminability measure relates to existing measures. We also review how the concept of discriminability has appeared in previous work.

\subsection{Quality Measures}

Most quality measures for visualization score the quality of a single view of a dataset. They are numerical functions whose arguments include an image or some visualization description, a dataset, and a task. In contrast, discriminability only makes sense if computed over a \textit{family} of datasets. Our discriminability measure scores the quality of an \textit{encoding} given an arbitrarily large data space. Despite this distinction in scope, many measures have been proposed that have features in common with our measure.

Behrisch et al.~\cite{Behrisch2018} showed that most quality measures in the visualization literature were not developed for evaluation purposes; instead, their purpose is to enable querying of visual patterns or automated recommendations. For example, assuming a user is looking to find clusters and there are too many possible views of a large dataset, the Hough Space measure for parallel coordinate plots \cite{Tatu2009} can be used to automatically find views where clusters are well-defined. Most measures are specific to task and/or encoding. There are measures for scatterplots~\cite{Wilkinson2005, Bertini2004}, node-link diagrams~\cite{Dunne2015}, line charts~\cite{Ryan2019}, and cartograms~\cite{Alam2015}, to name a few; and for class separation \cite{Tatu2009}, outlier detection~\cite{Johansson2009}, and change detection~\cite{Tu2007}.

A few measures that are more general in scope could be used to evaluate new encodings. The information-theoretic measures of Chen and J{\"{a}}nicke~\cite{Chen2010}, which include \textit{Visualization Capacity} and \textit{Visual Mapping Ratio}, attempt to quantify the intrinsic ``power'' of a visual encoding and its quality given a particular dataset. Similarly, the saliency measure proposed by J{\"{a}}nicke and Chen \cite{Janicke2010}, evaluates the match between the saliency of a visualization against the saliency of the underlying data. Tufte proposed the \textit{Data-ink Ratio}, a measure grounded on the principle of minimalism that penalizes visual embellishments~\cite{Tufte2001}. In addition, several measures were proposed for quantifying crowding, occlusion, or overplotting~\cite{Ellis2006b, Brath1997, Bertini2004}.

Unfortunately, we have not observed these measures being applied to the evaluation of new encodings. One reason may be the fact that none has been empirically validated; that is, the connections between measure and effectiveness were not established.

\subsection{Discriminability}

Discriminability has been studied in visualization as a property of visual channels---how many distinguishable values they can be divided into \cite{Munzner}---and as a property of encodings---how to bind visual values to data values so as to ensure differences in data values can be perceived well \cite{Demiralp2014, Demiralp2014b}. It has been extensively used in the design and evaluation of color encodings \cite{Szafir2018}, including sequential \cite{Brychtova2017,Liu2018} and categorical color palettes \cite{Brychtova2017,Lin2013, Gramazio2017}, and as a criterion for texture design \cite{Holten2006} and glyph design \cite{Yoghourdjian2018}. Rensink suggested that discriminability should be considered as one of the evaluation measures of a scientific framework for visualization~\cite{Rensink2014}.

The graphical inference framework~\cite{Wickham2010e}, which proposes plot ``line-ups'' as a method for evaluating the significance of visual discoveries, relies on encoding discriminability. Within this framework, Hofmann et al.~\cite{Hofmann2012} computed the ``power'' of competing visual encodings, which is the extent to which they enable the identification of the observed data in a line-up of plausible distractors. This power---which stems from statistical power---is, in essence, discriminability.

In vision science, discriminability is often measured with the computation of just noticeable differences (JND). The higher the JND, the lower the discriminability. The JND paradigm appears frequently in visualization research (e.g., \cite{bartram2011whisper,Szafir2018, Yang2019}).

\section{The Discriminability Criterion}

At the core of the discriminability criterion is the premise that visualizations are visual embeddings of data~\cite{Demiralp2014b}; as such, the notion of visualization quality is fundamentally tied to the preservation of structures in the data. An important consequence of the visual embedding model is that changes in data should always yield perceptible changes in the visual representation. This requirement is known as the \textit{principle of unambiguous data depiction}~\cite{Kindlmann2014} and, in theory, can be assessed with a discriminability test. Likewise, the principle of \textit{visual-data correspondence}, which states that changes in data should yield changes of equivalent magnitude in the visual representation, can in theory be verified in terms of discriminability.

Low discriminability indicates that changes in data values result in low visual change, implying that viewers may have trouble decoding values accurately (due to ambiguity). It can indicate common visual mapping problems, such as low utilization of the space, narrow encoding range, overplotting, and high clutter. Discriminability is one of the most fundamental quality dimensions in visualization because a difficulty to decode values may affect other tasks, such as outlier detection, estimation of means, and visual comparison.

It is also closely linked to visualization scalability. With large datasets, most visualizations that do not rely on aggregation or sampling \textit{saturate}. This saturation can be described as follows: a saturated plot causes a family of datasets to be mapped to identical or similar plots. Consider the common ``hairball graph'' example, where adding more nodes to a saturated graph does not change the appearance.  A discriminability measure can determine the saturation levels of a visualization at increasing data scales. The curve formed by discriminability over data scale can be understood as a description of visual scalability.

Unlike measures that stem from vision science (e.g., clutter, saliency), which are applicable to all images (including natural images), discriminability is a quality of the visual mapping, a visualization-native property. In particular, the level of clutter of a plot may say something about its visual quality, but it says nothing about the visualization method as an instrument to observe data.

For the scope of this paper, we propose the following definition of discriminability:

\begin{description}
  \item[Discriminability] Given a collection of datasets, the average perceptual distance between the corresponding visualizations.
\end{description}

Thus, a discriminability test comprises a data scope, which is given by the collection of datasets, a visual encoding method, and a visualization similarity (or distance) measure. The similarity measure can be either an empirical function---judgments are collected through user studies---or a computational measure that approximates the empirical judgments. Because our motivation is to discover cheap and fast methods for evaluating new encodings, we will examine here the use of a computational measure of similarity.

Alternatively, discriminability could be defined in terms of the average data distance needed to produce a just noticeable difference in the visualization. Or, given a seed dataset and corresponding visualization, the effort needed to produce a second dataset (beyond a certain data distance) that yields an ambiguous visualization. If an intelligent agent is trained to generate such ambiguity-inducing dataset pairs, the effort could be measured in terms of model complexity. This ambiguity induction is conceptually the same procedure proposed by Matejka and Fitzmaurice~\cite{Matejka2017} to generate wildly different scatterplots that have the same statistics.

\section{Measuring Structural Similarity}




In this section, we investigate in depth the possibility of a computational measure of similarity that can approximate human perceived similarity. With such a measure we could perform large-scale discriminability tests, involving not only variation in dataset size, but also in dataset distribution, entropy, etc.

The Structural Similarity Index (SSIM) was developed for quality assessment of compressed images~\cite{Wang2004}. Different than previous measures (e.g., mean squared error, and peak signal-to-noise ratio) that assumed that the perception of image quality depends on the visibility of errors, SSIM assumes that image quality depends on the preservation of structural information. As such, image quality can be quantified by a general measure of structural similarity between the original image and the compressed images. While the error-sensitivity paradigm tries to reproduce early-stage, low-level processing of the human visual system, such as thresholding informed by psychophysical experiments, the structural similarity paradigm tries to emulate the hypothesized \textit{function} of the overall human visual system. This function consists in probing the structures of observed objects.

The SSIM is defined as the weighted product of luminance similarity, contrast similarity, and structural similarity.
\begin{equation}\label{eq:ssim}
  SSIM(\boldsymbol{x},\boldsymbol{y}) = l(\boldsymbol{x},\boldsymbol{y})^\alpha c(\boldsymbol{x},\boldsymbol{y})^\beta s(\boldsymbol{x},\boldsymbol{y})^\gamma
\end{equation}
where $\boldsymbol{x \in R^D}$ and $\boldsymbol{y \in R^D}$ are vectors (of the same size) containing the grayscale pixel intensities of each image. The SSIM calculation normalizes the images with respect to luminance in the contrast similarity calculation, and then normalizes the images with respect to contrast in the structural similarity step. This way, the similarity components are made independent. We can think of Equation \ref{eq:ssim} as a pipeline (from left to right) where a feature is subtracted after it has been the subject of a similarity assessment.

\begin{figure*}[!t]
  \centering
  \includegraphics[width=\textwidth]{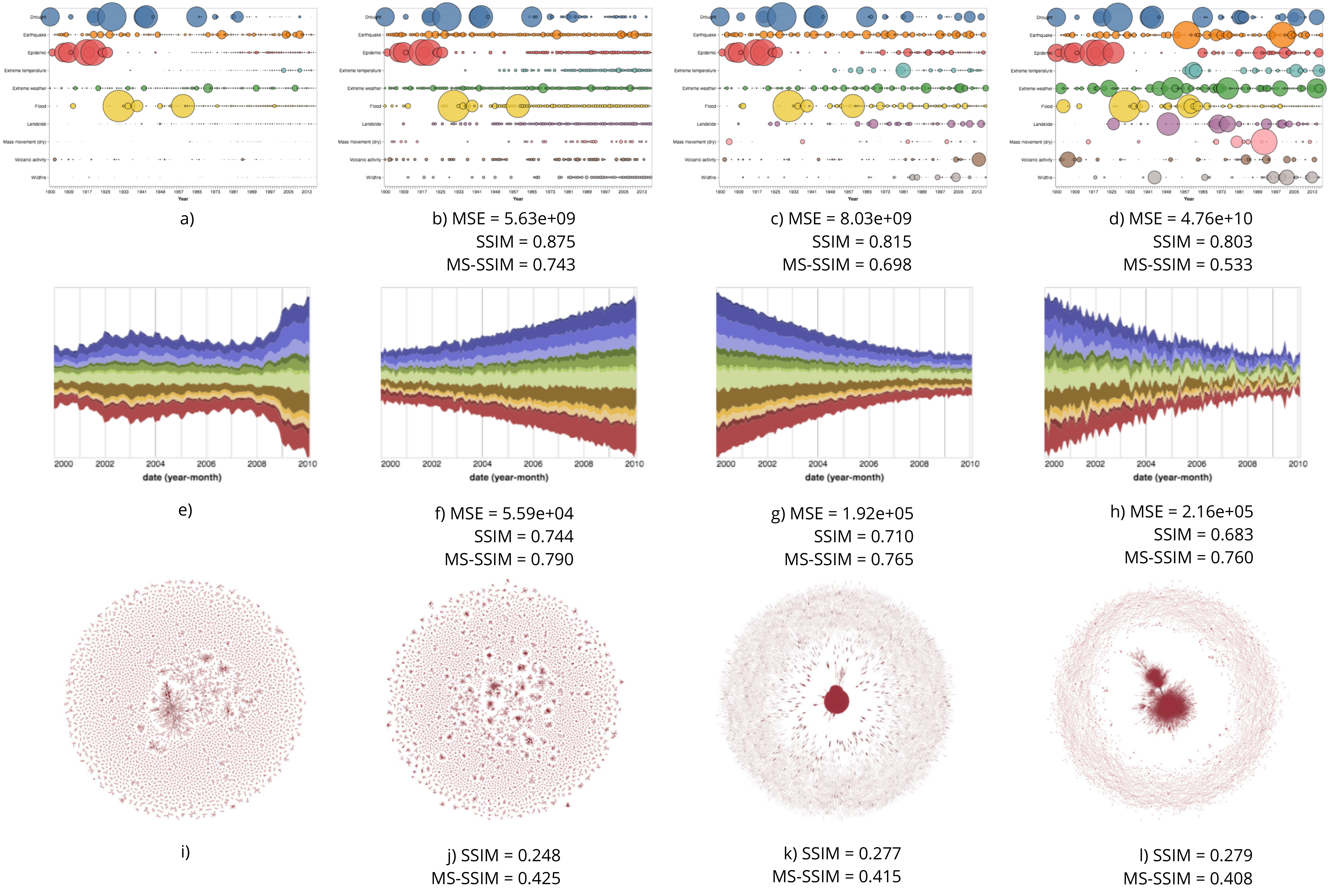}
  \caption[SSIM and MS-SSIM on visualization examples]{Data and image similarity measures: Mean-Squared Error (MSE), Structural Similarity Index (SSIM), and Multi-Scale SSIM (MS-SSIM). Leftmost images in each row are the references. Top: global deaths from natural disasters (Vega-lite gallery) and simulated perturbations. Middle: unemployment across industries (Vega-lite gallery) and simulated perturbations. Bottom: graphical models of passwords~\cite{Zheng2018}. MSE is inversely proportional to similarity. MS-SSIM weights: [0.1, 0.1, 0.1, 0.2, 0.5].}
  \label{fig:ssim-examples}
\end{figure*}


Luminance $\mu$ is the mean pixel intensity, and luminance similarity is defined as follows:
\begin{equation}
l(\boldsymbol{x},\boldsymbol{y})=\frac{2\mu_x\mu_y}{{\mu_x}^2+{\mu_y}^2}
\end{equation}
where $\boldsymbol{x}$ and $\boldsymbol{y}$ are vector representations of the images.  Contrast is estimated as the standard deviation of the pixel intensities. Note that the standard deviation $\sigma$ inherently subtracts the mean intensity (luminance) from the signal. Contrast similarity is defined analogously to luminance similarity:
%
%

%
\begin{equation}
c(\boldsymbol{x},\boldsymbol{y})=\frac{2\sigma_x\sigma_y}{{\sigma_x}^2+{\sigma_y}^2}
\end{equation}
Finally, the structural similarity function operates on the signal normalized by luminance and contrast: $(\boldsymbol{x} - \mu_x)/\sigma_x$. Readers familiar with machine learning will recognize this operation as \textit{standardization}, which yields a z-score. The structural similarity is the correlation (inner product) of these normalized vectors:
\begin{equation}
  s(\boldsymbol{\sigma_x}, \boldsymbol{\sigma_y})
  = \frac{1}{D-1} \sum_{i=1}^{D} \frac{(x_i - \mu_x)}{\sigma_x} \frac{(y_i - \mu_y)}{\sigma_y}.
\end{equation}

The SSIM is then computed in a local fashion (per pixel) with a 3x3 Gaussian window. This yields a similarity map over the image. The overall image similarity measure, a scalar value, is the mean similarity of this map:
\begin{equation}
\text{Mean-SSIM}(\boldsymbol{X},\boldsymbol{Y}) = \frac{1}{M} \sum_{j=1}^{M}\text{SSIM}(\boldsymbol{x}_j, \boldsymbol{y}_j)
\label{eq:mean-ssim}
\end{equation}
where $M$ is the number of Gaussian windows, $\boldsymbol{X}$ and $\boldsymbol{Y}$ are the images, and $\boldsymbol{x}_j$ and $\boldsymbol{y}_j$ are the image patches defined by each of the $M$ windows. When zero-padding is used, $M = D$. Despite the parent-child relation, the acronym SSIM usually refers to Mean-SSIM, and the distinction is rarely in effect. In this paper, we follow this convention. When the context suggests SSIM is a scalar value, it refers to the Mean-SSIM.


The SSIM is symmetrical, bounded, and has a unique maximum. The index lies in the interval $[-1,1]$ and a comparison between two identical images will always yield 1.

\subsection{Multiscale-SSIM}
\label{sec:ms-ssim}

Recall that the SSIM was created to measure the encoding quality of natural images, which depends on the impact of imperfections introduced by the encoding. Clearly, the perception of quality depends on the viewing distance, given that some imperfections are only noticeable at close inspection. In general, we can say that the perception of quality and similarity depends on the \textit{scale} of the image, which varies with viewing distance or image size.
Recognizing the challenges of assessing image quality at a single scale, Wang et al.~\cite{Wang2003} proposed Multi-Scale SSIM. This technique is a straightforward extension of SSIM where the contrast and structural similarities are computed at $K$ image scales. The original image is subject to low-pass filtering and downsampling by a factor of 2 in each of $K-1$ steps.
\begin{equation}
{MS{\text -}SSIM}(\boldsymbol{X},\boldsymbol{Y}) =
l(\boldsymbol{x},\boldsymbol{y})^\alpha
\prod_{i=1}^K c(\boldsymbol{x_i},\boldsymbol{y_i})^{\beta_i}
              s(\boldsymbol{x_i},\boldsymbol{y_i})^{\gamma_i}
\end{equation}

The weights indexed by $i$ are adjusted according to the desired relative importance of the scales to the similarity judgment. For simplicity, and following Wang et al. \cite{Wang2003}, we always set $\alpha=1$, and $\beta=\gamma=w_i$ within each scale:
\begin{equation}
  {MS{\text -}SSIM}(\boldsymbol{X},\boldsymbol{Y}) =
  l(\boldsymbol{x},\boldsymbol{y})
  \prod_{i=1}^K \Big( c(\boldsymbol{x_i},\boldsymbol{y_i})
                s(\boldsymbol{x_i},\boldsymbol{y_i}) \Big) ^ {w_i}
\end{equation}

Throughout this paper we will use vector notation to communicate the scale parameters; for instance, in the parameter array $W = [w_1, w_2, ..., w_n]$, $w_1$ is the weight on the \newtext{finest, detailed view}  (largest image), while $w_n$ is the weight on the \newtext{coarsest, distant view} (smallest image).

\subsection{Comparing SSIM and MS-SSIM}

To begin assessing the utility of SSIM as a measure of visualization similarity we designed a small sanity test. We chose two visualizations from the Vega-lite visualization gallery \cite{VegaLiteGallery}, a bubble chart and a stream chart, and produced data perturbations of different magnitudes. Then we measured the similarity between the visualizations of the perturbed data and the original visualization. These visualizations have encodings of different nature: point and area. We added also a third set of visualizations, which consists of plots of graphical models of password lists~\cite{Zheng2018}. They were chosen because they are dense representations that tend to form distinct shapes.

Figure \ref{fig:ssim-examples} shows the mean squared errors (MSE) computed on the dataset pairs, and both MS-SSIM and SSIM computed on the corresponding visualization pairs. The MSE summarizes the differences in values from one dataset to the other. In this experiment, it represents the baseline or true dataset difference. Most charts of unaggregated data where clutter is not an issue should allow us to recover, with some effort, the MSE between two datasets by mapping the visual marks back to data values and computing the measure. In fact, there are tools designed with the specific purpose of extracting data from existing visualizations \cite{Mendez2016, Harper2014}.

The SSIM produced similarity rankings that reflect the MSE rankings in the bubble chart and stream chart cases: larger SSIM values should correspond to lower MSE values. In the dense graph case the true data similarity is unknown (\newtext{as we only have the images}), so we will resort to a qualitative assessment. It is rather clear that two of the plots (k and l) feature a dense central region that forms a solid red blob, while the other two plots (i and j) feature a more well-distributed pattern. The output of the SSIM comparisons indicates that this notion is not captured by the measure; the graph that is perceived as most similar (j) to the reference (i) received the lowest SSIM score.

It is plausible that the similarity of plots is judged at different scales depending on the kind of plot. For instance, dense graphs form distinct global shapes that override local similarity comparisons. Other visualizations, such as scatterplots, may or may not form global shapes. When a global shape is not formed, the similarity judgment is done at a lower level, by scanning the scene in search of differences, a process that is well captured by the windowed calculation of SSIM.

MS-SSIM is built on the premise that \textit{viewing conditions} determine the right scale. We, instead, posit that at \textit{identical} viewing conditions the scale in which similarity judgments varies with the chart type. We customized the weights as following, so as to give more importance to \newtext{differences in coarser features:} 
[0.1, 0.1, 0.1, 0.2, 0.5]. The resulting scores (Figure \ref{fig:ssim-examples}) reflect the correct similarity ordering of the dense graphs. As a bonus, the MS-SSIM scores also comply with the correct data MSE ranks for the stream charts and bubble charts.


\subsection{Limitations}

\begin{figure}
  \includegraphics[width=\columnwidth]{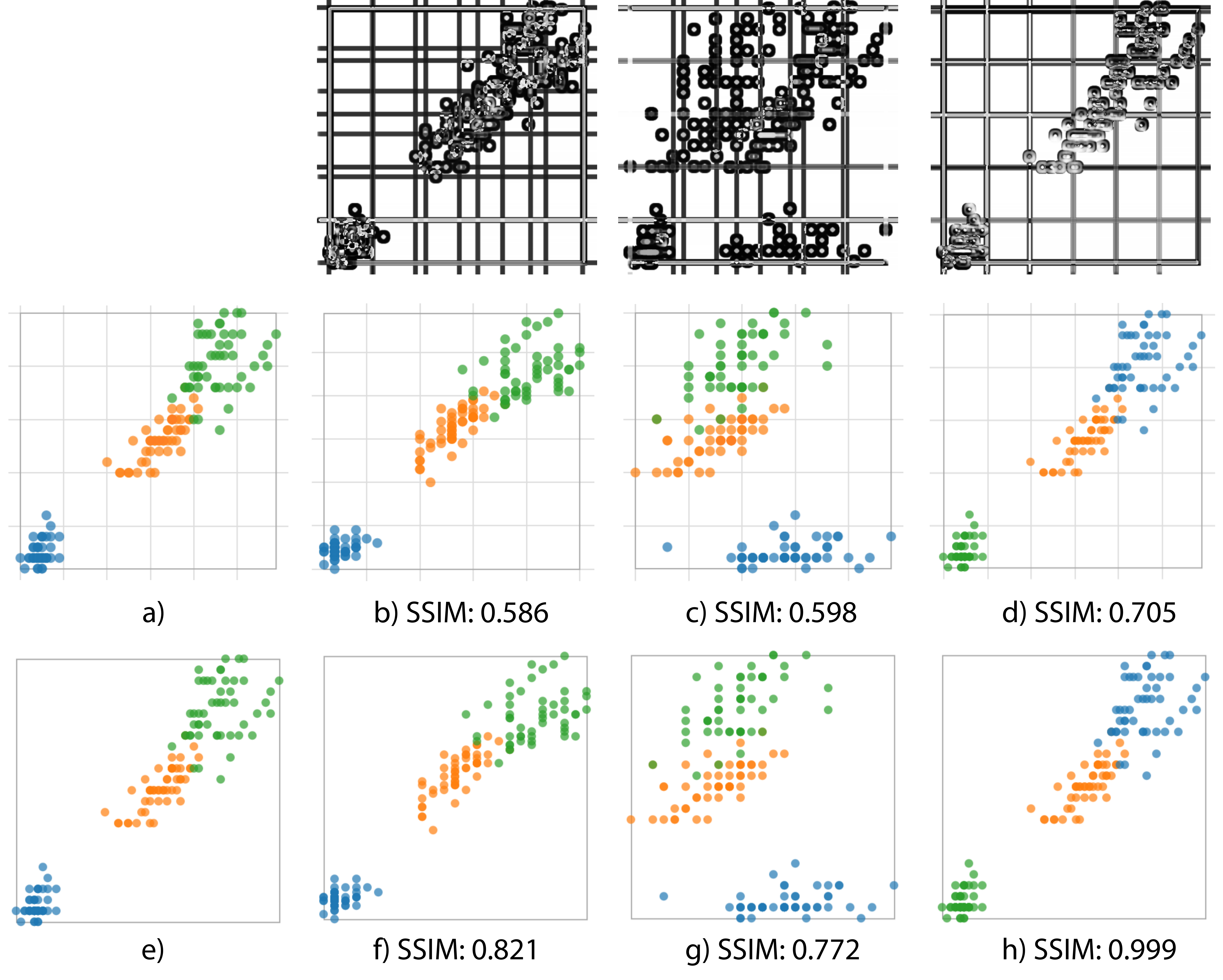}
  \caption[SSIM results on gridded plots]{The effect of grids on SSIM for scatterplots of the Iris dataset. (a-d): SSIM with grids. Top row: local SSIM values for these plots (brighter is more similar). (e-h): the same comparison without grids. Scores are relative to the leftmost plots in each row.}
  \vspace{-3mm}
  \label{fig:stress-tests-ssim-grids}
\end{figure}

Fundamental limitations arise when the SSIM is applied to data plots. In natural images every pixel counts towards a similarity judgment, although some extensions of the SSIM recognize that some regions matter more than others and attempt to weigh their importance based on saliency~\cite{Moorthy2009}, recognized objects~\cite{Ninassi2007}, and information theoretic measures~\cite{Wang2011}. In data plots, this characteristic manifests adversely as a hypersensitivity to visual accessories, such as grids and labels. Figure~\ref{fig:stress-tests-ssim-grids} (a-d) displays scatterplots of the Iris dataset that feature a grid. Note how the SSIM values do not correspond to the visual similarity of the plots. Upon close inspection we see that the grids, which are not consistently positioned, contribute disproportionately to the measurement.

In the context of the proposed use of the measure, the discriminability tests, the tester has control over the production of the images, so the hypersensitivity problem can be completely disregarded if we assume that for testing purposes, the plots are generated without grids, labels, and other accessories. \newtext{We also assume that the presence of these accessories would not change the result of a comparative evaluation. This is consistent with the findings of Bostock and Heer \cite{Heer2010}, who reported an effect of gridlines on effectiveness, but no interaction with chart type.}

Better measurements are achieved by simply turning the grid off (Figure~\ref{fig:stress-tests-ssim-grids} (e-h)). However, this figure illustrates a more complicated limitation. The scatterplot labelled (d) is a clone of (a) that had the color mapping inverted (blue and green swapped); therefore, (d) should not be judged identical to (a), as the SSIM value implies. The SSIM operates on grayscale images and it is not capable of capturing changes in hue.

The color limitation does not affect color encodings of numerical, continuous data attributes which employ color schemes that vary luminance and saturation. It affects more strongly visualizations that use nearly equiluminant categorical color palettes. In the next section, we propose a modification to SSIM that addresses its ``color blindness''.

\subsection{SSIM on YUV Color Space}

\begin{figure}
\includegraphics[width=\columnwidth]{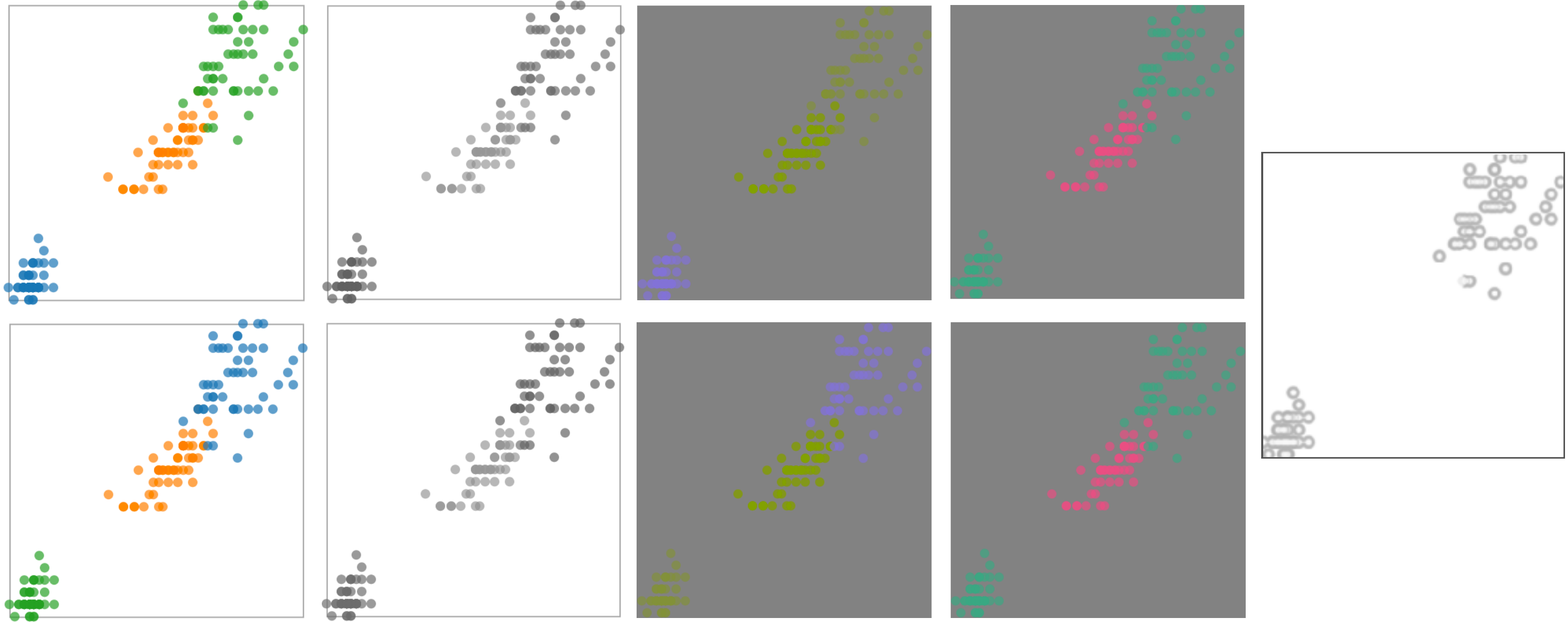}
\caption[SSIM on YUV color space]{SSIM applied on YUV image representations. Each row shows images in their original form, and decomposed into Y, U, and V channels of the YUV color space. On the right is the similarity map resulting from averaging the similarities computed on each channel independently. Note how the color difference in the original images appears in the final similarity map. \label{fig:yuv}}
\vspace{-3mm}
\end{figure}

Since in the next sections we will be investigating the discriminability of a broad set of encodings with various color mappings, it is important to establish a more general use of SSIM that can accommodate both categorical and continuous color mappings.

Our goal is to introduce \textit{some} sensitivity to color by using a color space where color components are represented independently from luminance. The YUV color space is well aligned with this goal, since it consists of a luminance component (Y), and two chrominance components (UV). Black and white images use only the Y component, so the original SSIM is equivalent to the computation on the Y channel.

We compute the SSIM on the YUV space by simply averaging the similarities computed in each color space component (Y, U, and V) independently. The computations on U and V can be interpreted as an assessment of the similarity existing in color structure. \newtext{Thus, the proposed calculation consists in the \textit{smoothing} of the original SSIM with values computed on color channels. Other color spaces that represent luminance or lightness independently are suitable, including the perceptually uniform ones; however, our calculations on individual components do not benefit from the perceptual uniformity property.}

In the pathological example depicted in Figure \ref{fig:yuv}, where two groups had their color swapped, this strategy is enough to prevent the visualizations from being scored identical. The standard SSIM similarity of 0.999 (Figure~\ref{fig:stress-tests-ssim-grids}(h)) fell to 0.968 using SSIM on YUV space. SSIM on YUV preserves SSIM's characteristic of being driven by spatial structure. Additional research is needed to determine the best approach if capturing color change is of primary importance. Given that it shows improvement over basic SSIM, we chose to work with SSIM on YUV when color is involved.

\section{Computing Plot Similarity}
\label{sec:empiricalvalidation}

In this section we compare MS-SSIM judgments with empirical similarity judgments. Our goal is to test whether a parameterization of MS-SSIM is capable of approximating empirical judgments for a certain visualization type. A positive result in this validation should indicate that other parameterizations can help us approximate judgments for other visualization types, assuming that the judgments will vary mostly with respect to scale and the use of color. If instead we find that no parameter set can approximate well empirical judgments, that should prompt discussion about what factors are involved in similarity perception of data plots. This applies in particular to spatial encodings.

For this analysis we chose the data collected by Pandey et al.~\cite{Pandey2016}, which consists of human similarity judgments (13 participants) for a set of 247 single-color scatterplots.
The similarity judgments were collected with a spatial arrangement interface
in which scatterplot thumbnails are displayed in an ``image carousel'' and can be dragged and dropped into a large, initially empty, canvas. Participants were instructed to
arrange the scatterplots into groups according to their similarity, explicitly mark the boundaries of each group, and finally, assign labels to them. Notably, they were told not to worry about within-group or between-group distances; that is, only group membership mattered.

\begin{figure*}
    \centering
    \subfloat[Empirical scatterplot clustering.]
    {
    \label{fig:stress-tests-pandey-clusters}
    \includegraphics[width=.48\linewidth]{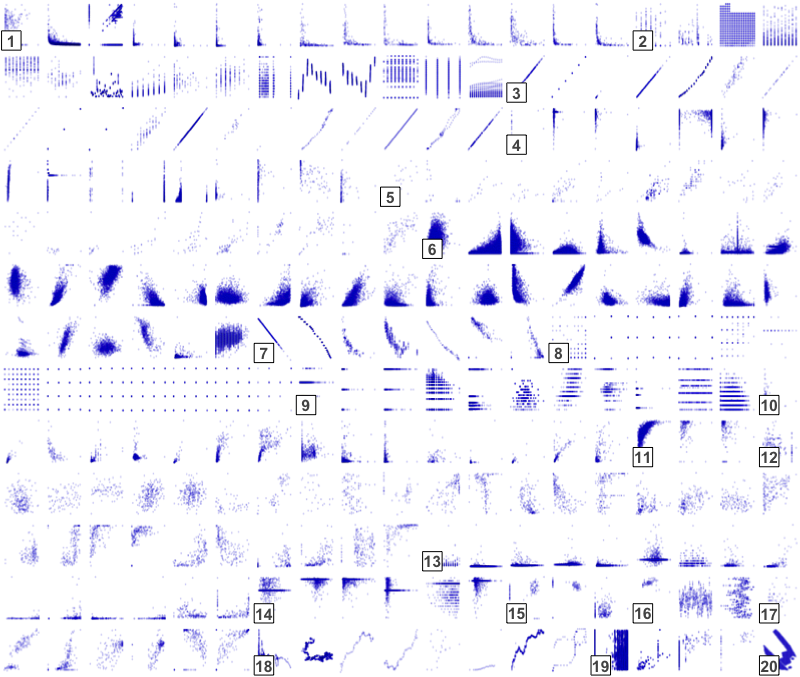}
    } \quad
    \subfloat[MS-SSIM scatterplot clustering.]
    {
    \label{fig:stress-tests-ssim-clusters}%
     \includegraphics[width=.48\linewidth]{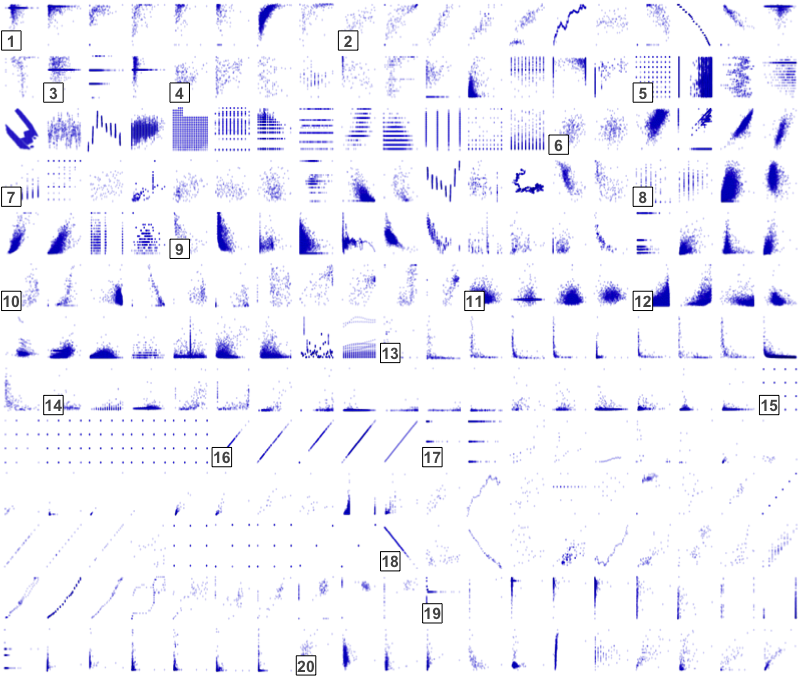}
    } \\

    \caption[Scatterplot clusterings \cite{Pandey2016}]{Empirical and MS-SSIM clusterings of the scatterplots from the study of Pandey et al. \cite{Pandey2016}. MS-SSIM parameters were tuned to the empirical data via gradient descent. Plots are ordered from upper left to lower right in a wrapped sequence. \label{fig:stress-tests-clusters}}
\end{figure*}


Pandey et al. \cite{Pandey2016} calculated the consensus distances for each pair of plots as the complement of their probability of co-occurrence averaged across participants:
\begin{equation}
  d_{i,j} = \frac{1}{N} \sum_{k=1}^N \Big(1 - \frac{c_{i,j}}{min(c_i, c_j)} \Big)_k
\end{equation}
where $N$ is the number of participants, $c_{i,j}$ is the number of clusters that contain both plots $i$ and $j$, and $c_i$ and $c_j$ are the number of clusters that contain the plots $i$ and $j$, respectively. Note that the interface allowed participants to assign plots to multiple groups. A hierarchical clustering of the plots based on the consensus perceptual distance matrix was calculated, and it is displayed in Figure \ref{fig:stress-tests-pandey-clusters}.

We compared MS-SSIM and empirical judgments using cluster quality measures, which are traditionally used to quantify the agreement between two independent label assignments on the same dataset. We selected the following measures, all of which assume the ground truth is known: adjusted mutual information (AMI), normalized mutual information (NMI), Rand index (RI), and adjusted Rand index (ARI).
All measures except RI assign values close or equal to 0 to random clusterings and assign 1 to perfect clustering (relative to the ground truth). Change adjusted measures (AMI and ARI) do not exhibit a dependency between the number of clusters and the number of samples; such dependency could boost the score of random clusterings that have many groups.

\newtext{In their experiments, Wang et al. \cite{Wang2003} found optimal MS-SSIM weight parameters for natural images (
0.04, 0.29, 0.30, 0.24, 0.13). In this section, MS-SSIM was set with five weight vectors manually chosen to represent different weight balancing strategies. As in the work of Wang et al., the vectors sum up to 1 and have components $<.5$. We compared these parameters with a weight vector tuned using gradient descent (the approach is described in detail in the next section). Table~\ref{tab:stress-tests-pandey-validation-results} presents the weight vectors ordered by importance on the finest scales.} The clustering method was 
hierarchical under the Ward agglomeration strategy, with even-height tree cuts that yielded 20 clusters (the same number of clusters in the ground truth, although none of the quality measures requires an equal number of clusters).

\begin{table}
\caption[Cluster quality measures]{Cluster quality measures for clusterings of 247 scatter plots based on MS-SSIM. The quality measures are relative to the clustering based on human similarity judgments reported by Pandey et al.~\cite{Pandey2016}. Each row corresponds to a parameter set ($w_1..w_5$). The parameters in the first row were obtained through gradient descent.}
\label{tab:stress-tests-pandey-validation-results}
\centering
\begin{tabular}{rrrrrrrrr}
\textbf{$w_1$} & \textbf{$w_2$} & \textbf{$w_3$} & \textbf{$w_4$} & \textbf{$w_5$} & \textbf{ARI} & \textbf{RI} & \textbf{AMI} & \textbf{NMI} \\ \midrule 

\textbf{0.32} & \textbf{0.73} & \textbf{0.82} & \textbf{1.00} & \textbf{1.00} & \textbf{0.20} & \textbf{0.90} & \textbf{0.35} & \textbf{0.51} \\
0.10 & 0.10 & 0.10 & 0.30 & 0.40 & 0.16 & 0.86 & 0.30 & 0.46 \\
0.10 & 0.20 & 0.20 & 0.20 & 0.30 & 0.13 & 0.83 & 0.25 & 0.42 \\
0.10 & 0.15 & 0.15 & 0.30 & 0.30 & 0.10 & 0.81 & 0.22 & 0.40 \\ 
0.20 & 0.20 & 0.20 & 0.20 & 0.20 & 0.13 & 0.81 & 0.24 & 0.42 \\
0.40 & 0.20 & 0.20 & 0.10 & 0.10 & 0.13 & 0.81 & 0.26 & 0.44 \\
 \bottomrule 
  \vspace{-2mm}
\end{tabular}
\end{table}

\begin{figure}[t]
  \centering
  \includegraphics[width=\linewidth]{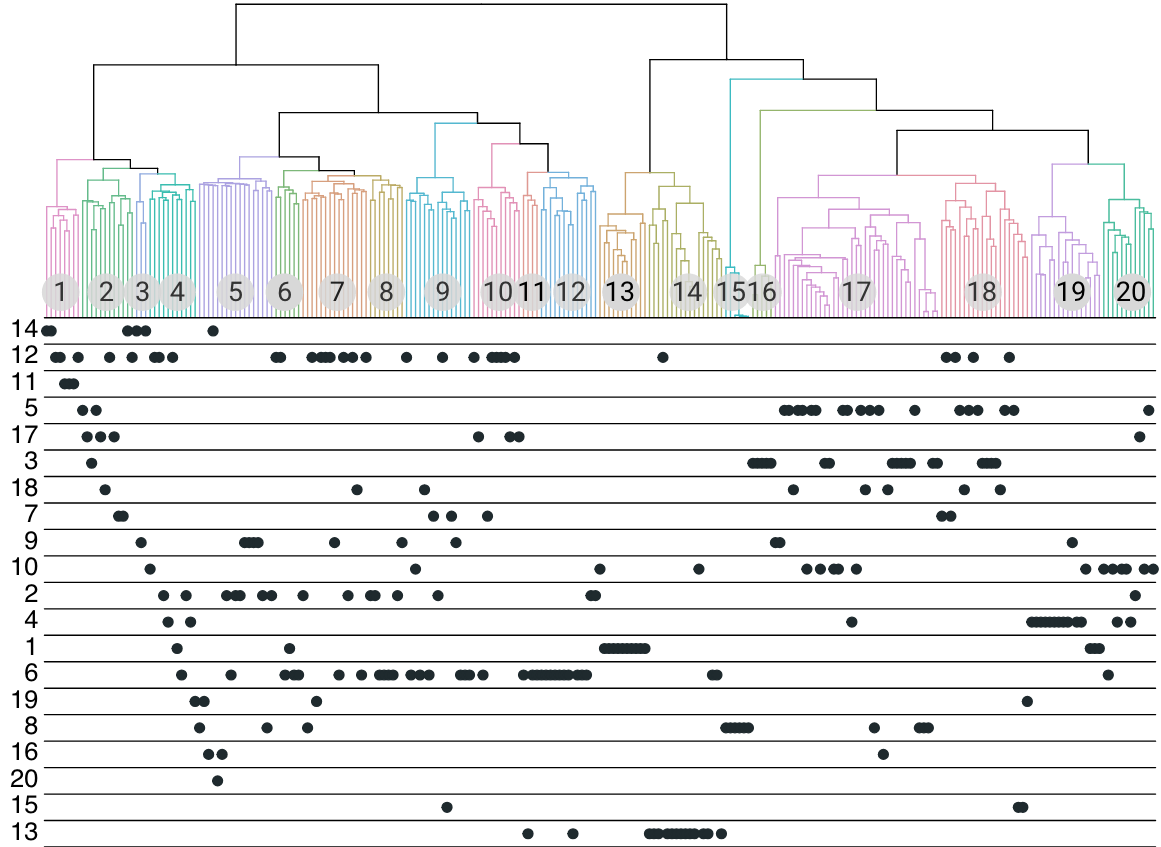}
  \caption[Dendrogram for MS-SSIM clustering]{A comparison of empirical and MS-SSIM clustering. The dendrogram represents the MS-SSIM clustering of scatterplots from Pandey et al.~\cite{Pandey2016}. Each row in the bottom represents an empirical cluster, with each dot representing a plot. Dots are aligned with the dendrogram, allowing us to observe how the empirical clusters are disrupted by the dendrogram arrangement. If the clusterings were identical, all dots in each row would be adjacent. Rows are ordered according to leftmost match with dendrogram. }
  \vspace{-2mm}
  \label{fig:stress-tests-ssim-dendrogram}
\end{figure}

The results (Table \ref{tab:stress-tests-pandey-validation-results}) show that the parameters found through gradient descent achieved the best fitness to the empirical clustering, as observed in all of the quality scores. The plot arrangement resulting from clustering with this best MS-SSIM parameter set is presented in Figure \ref{fig:stress-tests-ssim-clusters}, and the corresponding dendrogram in Figure \ref{fig:stress-tests-ssim-dendrogram}. The fitted parameters and the plot arrangement comparison tell us much about the protocol used to collect the empirical measurements.
First, the participants had only the chance of interacting with thumbnails, forcing them to make high-level perceptual judgments. This fact is expressed in the weights discovered with gradient descent, which clearly emphasize coarser judgments. Second, distances between and within clusters were not taken into account; as a consequence, the global structure in the empirical clusters is messy. The MS-SSIM clustering, in contrast, imposes a clear partition between dense and sparse plots (around cluster 13).


The cluster quality measures suggest a significant overlap between the clusterings, but far from full agreement. Some of the mismatch can be explained by the \textit{cognitive interaction problem} \cite{Wang2003}, by which different user goals can result in very different judgments.
 Participants were not instructed to cluster plots based on \textit{dataset similarity}. In a real-world scenario, analysts are making judgments about the data, with the visualization being a proxy. Some pairs of plots that bear some visual resemblance (in terms of shape) and are in the same empirical cluster, are unlikely to have been found similar if the question was about the underlying data. For instance, \includegraphics[height=1.5\fontcharht\font`\B]{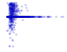} and \includegraphics[height=1.5\fontcharht\font`\B]{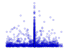} have both a T-like shape, but represent very different relationships between the data variables. We can attribute much of the difference between the clusterings to this misalignment of goals. Empirical cluster \#6, the one whose elements are spread the most across SSIM clusters, comprises elements with wildly distinct data patterns, but similar density (see Figure \ref{fig:stress-tests-clusters}). Density-based agglomeration is still present in the SSIM clustering, but divided according to the position of the point-cloud. Likewise, empirical cluster \#12 has plots with similar amount of ``ink'' but very different spatial arrangements; it is also divided in several pieces in the MS-SSIM clusterings.


\section{Tuning}

\newtext{This section explains in detail the MS-SSIM tuning procedure that we used to obtain the weights presented in the previous section. This procedure adjusts the scale weights so as to minimize the discrepancy between SSIM similarity and a set of empirical judgments.}
We used a vanilla stochastic numerical gradient descent algorithm, which,
at each iteration, evaluates the gradient of the loss function with respect to the current parameters, then updates the parameters in the directions that reduce the loss.

Let's define a dataset of images $x_i \in R^D$, and a similarity function $s : R^D \times R^D \rightarrow R^1$. With the multi-scale SSIM, $s$ has the following form:

\begin{equation}
  s(x_i, x_j) = \text{MS-SSIM}(x_i, x_j, W)
\end{equation}

The above equation can be read as the similarity of $x_i$ and $x_j$ given the vector of weights $W$, which determines the importance of each scale to the overall similarity score, as seen in Section \ref{sec:ms-ssim}. Next, let's define a binary function that takes an image triplet $(x_i, x_j, x_k)$ and decides whether $x_i$ is more similar to $x_j$ than $x_i$ is to $x_k$:

\begin{equation}
  f(x_i, x_j, x_k) = \mathbb{1}(s(x_i, x_j, W) \geq s(x_i, x_k, W))
\end{equation}

\newtext{This equation embodies a triplet matching task and enables the definition of a loss function for comparison of SSIM scores with a ground truth.
The ground truth data could be collected using triplet matching, triplet discrimination, spatial arrangement, or pairwise ratings on a Likert scale. }

The loss function is defined as follows, where $f_{ijk}$ is an abbreviation for $f(x_i, x_j, x_k, W)$, the SSIM binary label, and $Y_{ijk}$ is the ground truth label:

\begin{equation}
  L_{ijk}(W) = \sum_{f_{ijk} \ne Y_{ijk}} \Big(s(x_i, x_j, W) - s(x_i, x_k, W)\Big)^2 + R(W)
\end{equation}

The loss defined in the equation above is composed of two terms, the data loss and the regularization loss. The data loss is simply the squared difference between the similarity scores when they are wrong. For instance, if $s(x_i, x_j, W) = 0.8$, $s(x_i, x_k, W) = 0.6$, and the ground truth is $s(x_i, x_j, W) < s(x_i, x_k, W)$, that is, $Y_{ijk} = 0$, then the loss is $(0.2)^2$. The regularization loss (or penalty) is a function of the weights and embeds our preference for weights in a certain range. In this case, the weights need to be between 0 and 1. The regularization loss has the following form
\begin{equation}
    R(W) = \sum_{i=1}^{|W|} (W_i)^{\alpha-1}(1-W_i)^{\alpha-1}
\end{equation}
where $\alpha$ is a parameter that controls the steepness of the penalty as the values approach 0 or 1.

\section{Discriminability and Effectiveness}

\begin{table}
\caption[The tasks of Kim and Heer \cite{Kim2018}]{Kim and Heer's experiment was divided into four tasks. $Q_1$ is a continuous variable. }
\label{tab:stress-tests-kim-tasks}
\begin{tabular}{ll}
\toprule
Read value & What is the $Q_1$ of the data point A? \\
Compare value & Which data point has more/less $Q_1$? \\
Find maximum & \begin{tabular}[c]{@{}l@{}}Which state has the data point with the \\ highest $Q_1$?\end{tabular} \\
Compare averages & \begin{tabular}[c]{@{}l@{}}Considering all data points for the State, \\ which of the following two States has greater \\ average $Q_1$?\end{tabular} \\
\bottomrule
\end{tabular}
\vspace{-5mm}
\end{table}

The validation of the MS-SSIM against an empirical study of scatterplot similarity was useful for understanding the extent to which we can expect human similarity judgments to match MS-SSIM scores, but it did not shed light on the usefulness of discriminability as a quality criterion. We don't know if discriminability scores derived from similarities have any relationship to the effectiveness of visualizations. In this section, we seek to fill this gap.

There are a few empirical studies of the effectiveness of visualization encodings. We will base our investigation on the most recent of these studies, which has all materials publicly available~\cite{Kim2018}. As a plus, this study focused on the effect of data scale and distribution on performance, so it aligns with our interest in scalability. Kim and Heer~\cite{Kim2018} tested the effectiveness of twelve trivariate encodings, described here in the format $Q_1$\_$Q_2$\_$N$ where $Q_1$ and $Q_2$ are numerical, continuous variables, and $N$ is a categorical variable.

 The data consists of 2016 U.S. monthly weather measurements, published as part of the Global Historical Climatology Network-Daily Database (GHCN)~\cite{Menne2012}, and contains the categorical variables State and Month, and the following numerical variables: Maximum Temperature, Minimum Temperature, Average Wind Speed, Wind Direction, Strongest Gust Speed, Precipitation, Snowfall, and Snow Depth.

The stimuli of that experiment was produced by sampling from GHCN and it was divided into 24 experimental conditions that result from the crossing of the following factors: Cardinality (3, 10, 20), where cardinality is the number of categories $N$, \#/Category (3, 30), $Entropy_{Q_1}$ (Low, High), and $Entropy_{Q_2}$ (Low, High). The specific variables $Q_1$ and $Q_2$ were not factors; thus, they vary randomly across stimuli. $N$ is always a derived variable resulting from the conflation of State and Month (as in TX-03), although in the stimuli it appears simply as State (participants were not exposed to Month).

Study participants were asked to perform tasks that involved questions about $Q_1$. The tasks were of the following types: Value tasks, further split into Read Value and Compare Value; and Summary tasks, further split into Find Maximum and Compare Averages. Table~\ref{tab:stress-tests-kim-tasks} lists the question templates for each task. Error rates and completion times were measured, and rankings of encodings were created based on the error rates.

The results of this experiment reveal that the effect of encoding on error rates depends on the task and on the various factors manipulated in the experiment; therefore, a different ranking of encodings is created within each task group and factor level. Furthermore, the differences in error rate and completion time for the encodings are not always statistically significant; for instance, in summary tasks involving datasets with three and ten categories, the ten best ranked encodings did not score significantly different error rates.

\subsection{Measuring Discriminability}

\begin{figure}
  \includegraphics[width=\columnwidth]{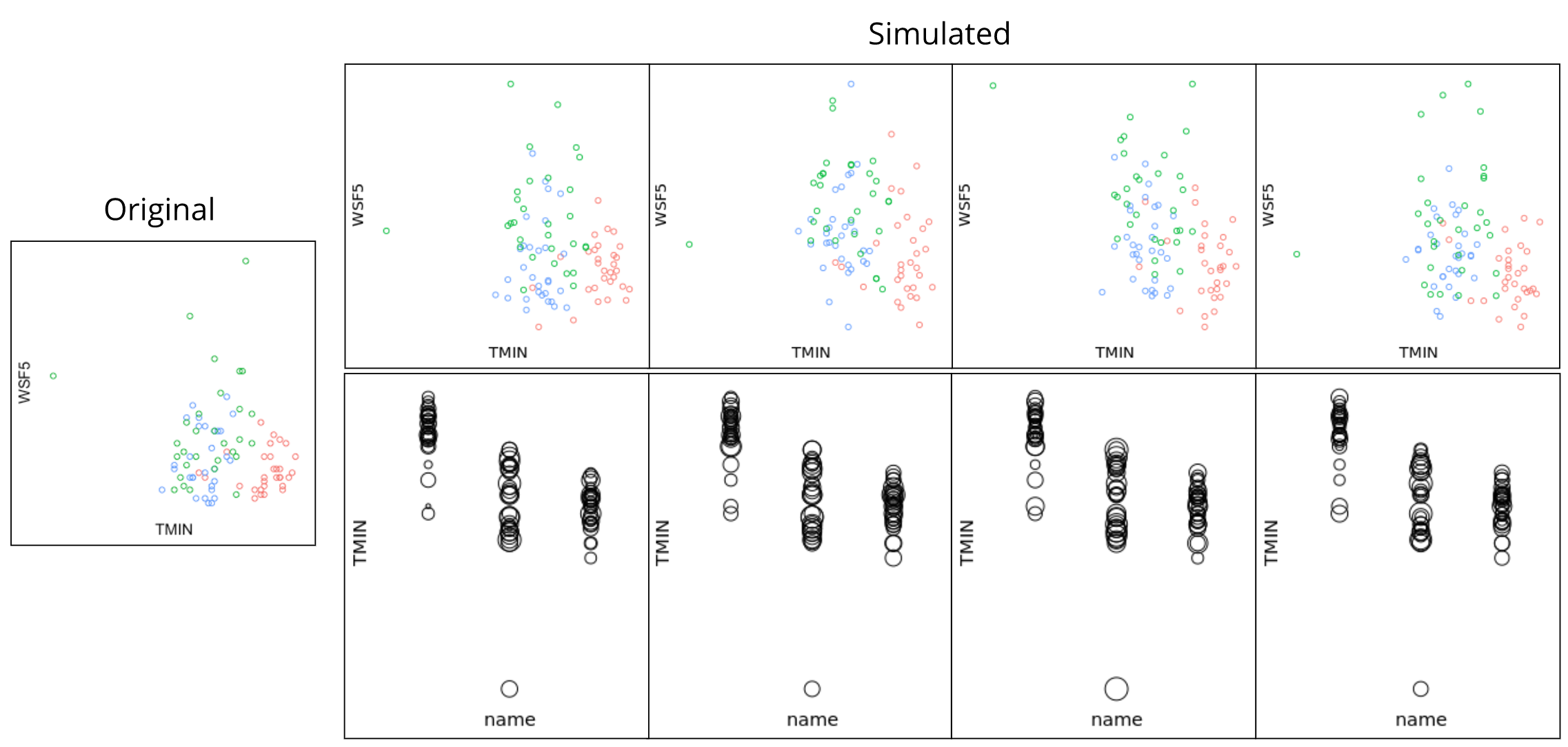}
  \caption[Global discriminability test sample]{Images generated for the global discriminability test. Left: Original plot used by Kim and Heer. Right: Plots depicting variations of the original data, resulting from sampling from statistical models fitted to Kim and Heer's data. Only the question variable $Q_1$ (WSF5 in this example) is simulated. The simulated data is depicted using position encoding (\textit{y\_x\_color}), and size encoding (\textit{size\_y\_x}) for $Q_1$. }
    \vspace{-2mm}
  \label{fig:stress-tests-kim-simulation-example}
\end{figure}

We conducted two benchmarks. The first is a \textit{global} discriminability test, of the kind someone would run without a specific task in mind. It generates a variety of datasets then computes the average similarity across visualizations of these datasets for each encoding being considered. In essence, it measures the sensitivity of each encoding, or how much \textit{overall} visual change we can expect of each encoding, on average. The link to effectiveness is in the assumption that the less sensitive an encoding, the harder it is to decode information: reading and comparing values is more difficult when the visual range is narrow.

\begin{figure}

    \centering
    \subfloat[]
    {\includegraphics[width=.25\linewidth]{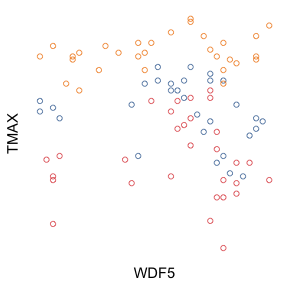}} 
    \subfloat[]
    {\includegraphics[width=.25\linewidth]{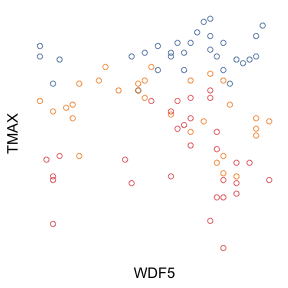}} 
    \subfloat[]
    {\includegraphics[width=.25\linewidth]{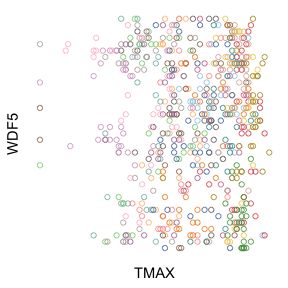}}
    \subfloat[]
    {\includegraphics[width=.25\linewidth]{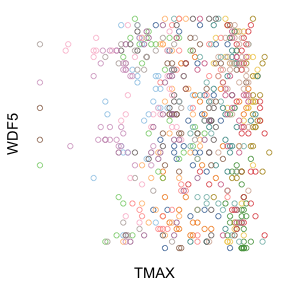}} \\

    \caption[Images for local discriminability test]{Pairs of colored scatterplots (\textit{y\_x\_color}) with y values swapped between two categories. a) and b) have 3 categories in total, while c) and d) have 30 categories. These pairs (a,b) and (c,d) are used to measure the visual discriminability of two categories (other categories fixed) along one variable. }\label{fig:local_discriminability_setup}
\end{figure}

The second experiment is task-specific. Kim and Heer's rankings for summary tasks (mean comparison and find maximum) are somewhat different than the rankings for value tasks. In the mean comparison tasks, participants are instructed to select the state with the highest mean out of only two options. It is safe to assume that in these tasks what matters is how easily people can segregate the values of the two states in question and compare their values. So we devised a scheme to test \textit{local} discriminability in Benchmark 2.

\begin{figure*}[t]
  \includegraphics[width=\linewidth]{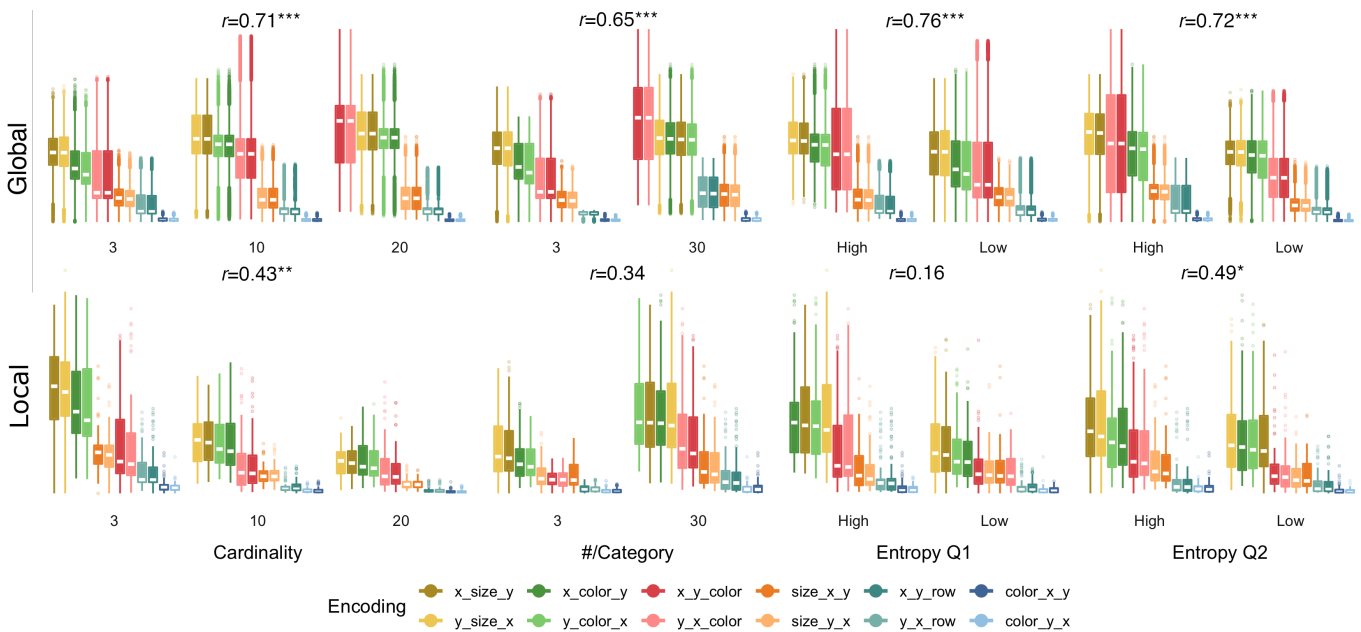}
  \caption{Global and Local discriminability scores computed with MS-SSIM ($W=[1,1,1,1,1]$), aggregated by data attributes. For each dimension, we computed the Pearson correlation coefficient against the empirical accuracy measured by Kim and Heer~\cite{Kim2018}, using the mean values for each encoding. The significance codes correspond to the null hypothesis that correlation is 0. Significance: 0 `***' 0.001 `**' 0.01 `*' 0.05 `.' 0.1 ` ' 1}
  \vspace{-5mm}
  \label{fig:discriminability_boxplots}
\end{figure*}

\subsection{\newtext{Benchmark} 1 - Global Discriminability}

Kim and Heer's experiment is structured as follows: 8 different datasets were sampled from the GHCN records for each combination of factors \textit{cardinality} $\times$ \textit{\#/category} $\times$ $entropy_{Q_1}$ $\times$ $entropy_{Q_2}$ $\times$ \textit{encoding}. That is, within each condition, each encoding was tested with a different collection of datasets, all with similar characteristics (dictated by the experimental condition). The datasets vary randomly in $Q_1$, $Q_2$, and the specific data points and states that the questions center on, in order to avoid a combinatorial explosion of conditions.  In the discriminability tests, we prioritized symmetry by testing all encodings within a given experimental condition on the \textit{same} datasets.
Furthermore, $Q_1$ and $Q_2$ were not varied randomly; instead, they were a factor in the experiment (between-encodings).
These changes were made because the scale of the test is not a problem here, so we can test every possible cross between $Q_1$, $Q_2$, and the rest of the factors. In summary, we created 20 datasets by simulation for every combination of factors \textit{cardinality} $\times$ \textit{\#/category} $\times$ $entropy_{Q_1}$ $\times$ $entropy_{Q_2}$ $\times$ $Q_1$ $\times$ $Q_2$.

In order to simulate data that are similar to the data used by Kim and Heer~\cite{Kim2018}, we sampled values from generalized linear models (GLMs) fitted to the GHCN data. The simulation consisted in randomly drawing a dataset that matched the given experimental condition, then replacing its $Q_1$ values by values sampled from the model. The replacement step was repeated 20 times.
The GLMs were fitted as follows. Given a condition, all records in Kim and Heer's data that match $Q_1$ were collected. Then a GLM was fitted to these records with $Q_1$ as the response variable and State as the covariate. Since all datasets have low correlation, $Q_2$ was omitted from the model; thus, the GLMs simply learn one distribution for each state. Figure \ref{fig:stress-tests-kim-simulation-example} shows a reference dataset and simulated datasets visualized with two different encodings.

\begin{figure*}[t]
    \centering
    \includegraphics[width=.95\textwidth]{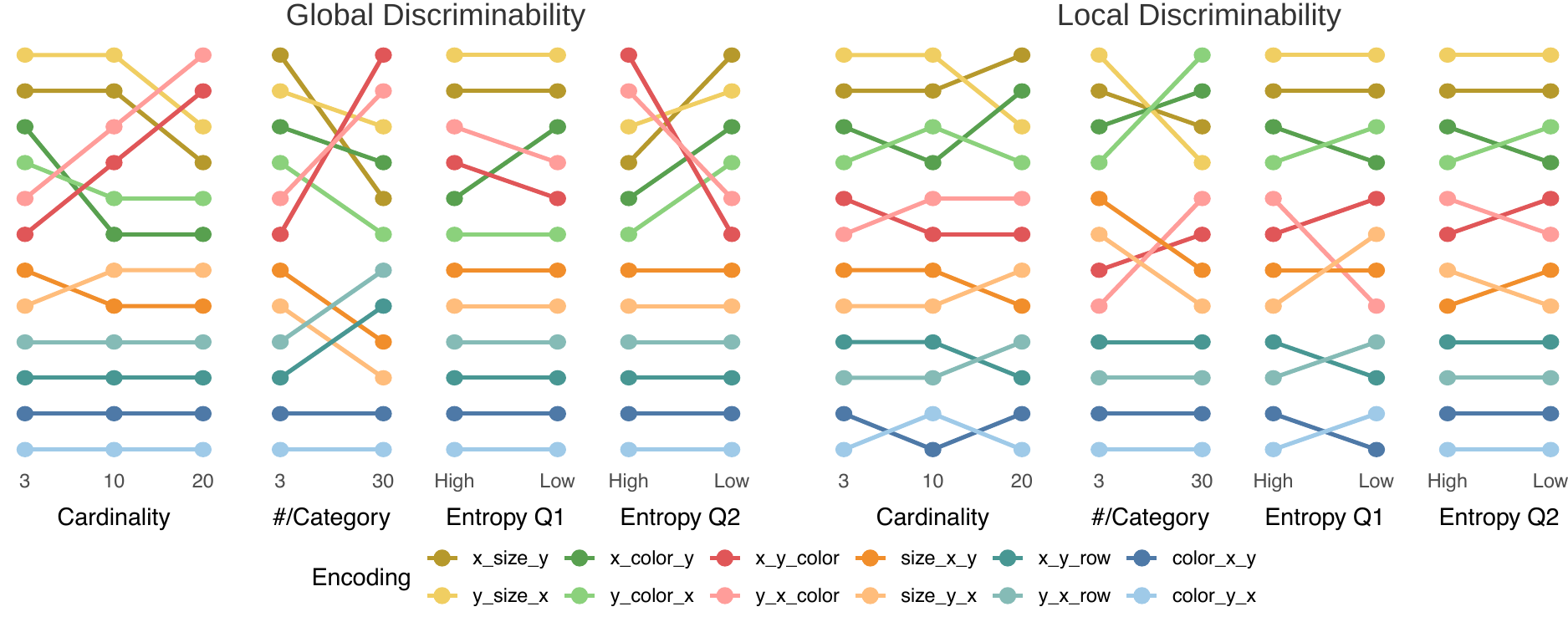}
    \caption{Discriminability rankings of encodings (divided by data property) derived from discriminability scores.}
      \vspace{-4mm}
      \label{fig:discriminability_rankings}
\end{figure*}

For each encoding, pairwise similarity judgments were computed with the MS-SSIM on the YUV representations of the images. We used a uniform parameter vector ($W=[1,1,1,1,1]$), \newtext{which embodies the ``naive'' hypothesis that all scales contribute evenly to similarity judgments. We assume that, \textit{a priori}, the parameters fitted to the scatterplots in Section \ref{sec:empiricalvalidation} are not applicable because the tasks and the presentation of the stimuli are quite different.} We transformed these similarities into distances by computing $d(x,y) = (1 - \text{MS-SSIM}(x,y))/2$. Each tuple (\textit{cardinality}, \textit{\#/category}, $entropy_{Q_1}$, $entropy_{Q_2}$, $Q_1$, $Q_2$, \textit{encoding}) yields one discriminability score computed as the mean pairwise distance over 20 images.  These scores are then aggregated to produce scores per factor level, used in the rankings of encodings. We compared the results of this experiment with the effectiveness data for value tasks.

\subsection{\newtext{Benchmark} 2 - Local Discriminability}

As mentioned earlier, we cannot expect a general test as the one presented in Benchmark 1 to explain accurately the effectiveness of a task that requires the comparison of two sections of a visualization, because that experiment evaluated global discriminability.

In order to test the discriminability of the visual representations of the two categories \textit{within the context of the whole plot} we devised the following testing scheme. Given a plot, a subset of two categories, and the variable $Q_1$ subject to the mean comparison, a second plot is generated where the values of $Q_1$ are swapped between the two categories. The values for $Q_2$ in both categories remain fixed, as well as all data points in all other categories. The similarity is then computed on this pair of images, effectively measuring the visual similarity of the two groups of data points in the context of the rest of the data. Figure \ref{fig:local_discriminability_setup} illustrates the scheme.

This test did not employ statistical simulation. We modified the same datasets that served as stimuli in Kim and Heer's experiment, which had 2,304 mean comparison tasks. In our experiment, each of these datasets was modified once, resulting in 4,608 datasets. Discriminability was calculated as the average distance (as described in Benchmark 1) between source and modified datasets. We assigned the same weights to all scales ($W=[1,1,1,1,1]$). We compared the results of this experiment with the effectiveness data for summary tasks.

\subsection{Results and Discussion}

\newtext{We used Pearson correlation to compare encoding discriminability scores with empirical effectiveness (task accuracy) within each study factor. The analysis of statistically significant differences within ranks is left for future work.} The global discriminability experiment produced encoding discriminability scores that are highly correlated to empirical effectiveness in value tasks (Figure~\ref{fig:discriminability_boxplots}).
The derived rankings in Figure~\ref{fig:discriminability_rankings} are nearly identical to the value task rankings of Kim and Heer, with spatial encodings exhibiting higher discriminability compared to encodings that rely on size and color, the exception being the spatial encoding \textit{x\_y\_row}, which displays categories in different axes. \newtext{The full details of the benchmark scores are provided as supplementary materials.}

The local discriminability experiment produced encoding discriminability scores that are mildly correlated to empirical effectiveness in summary tasks (Figure~\ref{fig:discriminability_boxplots}). In particular, discriminability scores did not account for the radical drop in accuracy (relative to value tasks) of the encoding \textit{x\_y\_color} (multi-class scatterplot). This drop is due to the difficulty humans have in separating colors in displays with many colors. Neither the sharp increase in the effectiveness of \textit{size\_x\_y} was observed. But curiously, the local discriminability of these encodings did suffer changes (relative to global discriminability) in the same directions observed in the empirical data, even though the magnitude was not equivalent. The failure to account for the difficulty in separating colored groups in plots with many groups~\cite{Haroz2012} is the main limitation of MS-SSIM as a measure of discriminability.  Future efforts should concentrate on measures that account for complex perceptual effects.

The correspondence between the rankings suggest that the effectiveness of the encodings is, to a large extent, driven by encoding discriminability. The discriminability within six out of eight factors had correlation with effectiveness higher than .4. However, fine grained changes in Kim and Heer's rankings due to data entropy and scale were not matched by the discriminability rankings. This suggests that discriminability cannot fully explain the rankings. This is to be expected, since other factors are known to influence people's judgments. Among these factors are saliency and distortions in the perception of brightness, contrast, length, and area (as described by Steven's law).

MS-SSIM can be interpreted as an inverse measure of the strength of the visual difference generated by a visual encoding. The successive downsampling steps measure the preservation of the difference at increasing viewing or judgment distance. The utility of multiscale representation in visualization was first explored by Wattenberg and Fisher~\cite{Wattenberg2004}. Moreover, the rankings of encodings suggest that the MS-SSIM seems to capture the notion of display space utilization, which Chen and J{\"{a}}nicke~\cite{Chen2009} define in information-theory terms. Encodings that tend to result in the least amount of blank space have higher discriminability.

\section{Final Remarks}

Computational measures of quality offer a scalable, low-cost, alternative to experiments with human participants. Discriminability, as a fundamental dimension of visualization quality, should be at the bottom of a stack of quality criteria for visual encoding evaluation. Future research should investigate what other criteria should compose this evaluation stack.
Coupled with data simulation, discriminability tests enforce data characterization. Designers and researchers are required to document the data parameters and boundaries wherein proposed encodings are expected to produce high quality plots. This practice strengthens statements of the generality of research contributions and helps other researchers identify opportunities for new research.

Discriminability scores are tools to verify the principles of visual-data correspondence and unambiguous data depiction. \newtext{In Kim and Heer's tasks there is little downside in perceiving a small data change as large. In other scenarios, proportionality is critical; for instance, a medical researcher examining effect size on a clinical experiment. In the future, discriminability scores may allow researchers to obtain an objective measure of the bias used in data communication. For instance, a scientific journal could create a standard maximum discriminability score for visualizations, in order to prevent exaggeration of effects.} Our work also opens new directions for the study of ambiguity in visual encodings, which constitutes an overlooked source of uncertainty in visual information analysis.

Our exploration of the MS-SSIM brought to surface the role of perceptual scale on similarity judgments. The scale where a visualization is read---the level of detail considered---impacts people's perception of similarity. Currently, we know of no comprehensive studies that address this issue. Meanwhile, at least one visual analysis protocol~\cite{Wickham2010e} relies on accurate readings of similarity. Verifying the hypothesis that different encodings afford similarity judgments at different scales is a topic for future research. \newtext{If this hypothesis is confirmed, MS-SSIM could help us 
discover and build a catalogue of these parameters for every chart type, in the same way that Steven's law has coefficients for stimulus types and Fitts' law has parameters that vary with device. This would require the design of a solid study protocol to collect data, and the fitting could be done as in Section 5. Informed by the appropriate weights, engineers could test the discriminability of a visualization on a dataset collection, and designers could test new designs by using weights fitted to visualizations that approximate the new design.}

\newtext{Our modifications to MS-SSIM add only some sensitivity to color. The proposed measure still embodies Wang's hypothesis that image similarity depends on preservation of \textit{spatial} structure.  Interestingly, the correlation with Kim and Heer's data is high despite our measure's lack of a sophisticated handling of color. So it is plausible that color is indeed less important to visualization discriminability (as it is the case with perceived image quality, Wang's problem). The results are aligned with the consensus that the color channel allows the encoding of fewer distinct values \cite{Munzner}, which explains at least the poor performance of color encodings in the ``Read Values" task.}

\section{Conclusions}

In this paper, we introduced a general method for computing visual encoding discriminability that requires only a collection of datasets and the corresponding rendered visualizations. While discriminability has been a quality criterion in visualization for a long time, it has been mainly confined to theoretical discussions. This work constitutes the first methodical application of the discriminability criterion to the evaluation of visualization encodings.

We examined the suitability of SSIM and MS-SSIM for scoring plot similarity, and revealed limitations related to over-sensitivity to visual accessories (e.g., grids) and failure to capture differences in hue. To overcome these limitations, we proposed modifications that achieved satisfactory results.
We demonstrated that a parameterization of the MS-SSIM can be found via gradient descent that achieves significant overlap with empirical plot similarity judgments. Most importantly, we devised a method for calculating encoding discriminability using the MS-SSIM and established a link between discriminability and task accuracy for a collection of basic encodings. We found that discriminability correlates with accuracy, especially for tasks that involve reading values of individual data points.

In conclusion, our results suggest that the MS-SSIM is useful for approximating plot similarity, and that discriminability scores based on MS-SSIM are associated with effectiveness. We recommend these scores to be used in early stages of visual encoding evaluation, as a low-cost computational measure of quality prior to committing to costly user studies.

 \acknowledgments{We acknowledge the support of the Natural Sciences and Engineering Research Council of Canada (NSERC) and Funda\c{c}\~{a}o CAPES (9078-13-4/Ci\^{e}ncia sem Fronteiras).}

\bibliographystyle{abbrv-doi}

\bibliography{refs}
\end{document}